\documentclass{aa}
\usepackage{latexsym, epsfig}
\usepackage{graphicx}
\usepackage{natbib}
\usepackage{times}
\usepackage{url}
\def\kms{\,{\rm km\,s}^{-1}}

\begin{document}

\title{The nature of turbulence in OMC1 at the scale of star formation:
  observations and simulations\thanks{Based on observations obtained at the
    Canada-France-Hawaii Telescope (CFHT) which is operated by the
    National Research Council of Canada, the Institut National des
    Science de l'Univers of the Centre National de la Recherche
    Scientifique of France, and the University of Hawaii.}}

\author{M. Gustafsson\inst{1}
\and A. Brandenburg\inst{2} \and J.\ L. Lemaire\inst{3} \and D. Field \inst{1}}
\offprints{D.Field}

\institute{Department of Physics and Astronomy, University of Aarhus,
  DK-8000 Aarhus C, Denmark \and NORDITA, Blegdamsvej 17, DK-2100
  Copenhagen $\emptyset$, Denmark \and Observatoire de Paris \& Universit\'e de Cergy-Pontoise, LERMA \& UMR
8112 du CNRS, 92195 Meudon, France}

\date{For Main Journal, Diffuse Matter in Space. Received:  Accepted:} 

\abstract{
}{
To study turbulence in the Orion Molecular Cloud (OMC1) by comparing
observed and simulated characteristics of the gas motions.  
}{
 Using a dataset of vibrationally excited H$_2$ emission in OMC1
 containing radial velocity and 
  brightness which covers scales from 70\,AU to 30000\,AU, we present the
  transversal structure functions and the scaling of the structure
  functions with their order. 
These are compared with the predictions of two-dimensional projections of simulations of
supersonic hydrodynamic turbulence. 
}{
The structure functions of OMC1 are not
  well represented by power laws, but show clear deviations below
  2000\,AU. However, using the technique of extended self-similarity,
  power laws are recovered at scales down to 160\,AU. The scaling of the
  higher order structure functions with order deviates from the standard scaling
  for supersonic turbulence.
  This is explained as a selection effect of preferentially observing
  the shocked part of the gas and the scaling can be reproduced using
  line-of-sight integrated velocity data from subsets of supersonic
  turbulence simulations.
  These subsets select regions of strong flow convergence and high
  density associated with shock structure. Deviations of the
  structure functions in OMC1 from power laws cannot however be
  reproduced in simulations and remains an outstanding issue.  
}{}
\keywords{ISM: individual objects: OMC1 - ISM: kinematics and dynamics
  -ISM: molecules - shock waves - turbulence - hydrodynamics}

\maketitle

\markboth{M. Gustafsson et al: }{}

\section{Introduction}
Turbulence plays a central role in star-forming molecular
clouds, acting both to support the clouds globally and to
create local clumps and density enhancements that can undergo gravitational
collapse. Simulations have shown that this latter process, known as turbulent
fragmentation, may directly determine
the initial mass function (IMF). Insight into the effects of
turbulence on molecular clouds is thus essential for understanding the
mechanisms of star formation. Such insight can only be gained by a
close interplay between
observations and simulations.

The characterization of turbulence may be achieved by
statistical methods. Several techniques, such as the size-linewidth
relation \citep{larson1981,goodman1998,ossenkopf02,gustafsson2005}, probability distribution functions
\citep{falgarone1990,falgarone1994,miesch1999,ossenkopf02,pety2003,gustafsson2005}, structure functions \citep{falgarone1990,miesch1994,ossenkopf02,gustafsson2005} and
$\Delta$-variance \citep{bensch2001,ossenkopf02}, have previously been used to
characterize the 
structure of brightness or velocity in molecular clouds.
Comparisons between observations and models have earlier been made by
\cite{falgarone1994,falgarone1995,lis1998,joulain1998,padoan1998,padoan1999,pety2000,klessen2000,ossenkopf02,padoan2003,gustafsson2005};
see also the review by \cite{elmegreen2004}. These
earlier comparisons \citep[save those in][]{gustafsson2005} are based on CO observations, tracing relatively
cool and low density gas. Data are limited in spatial resolution and
can only be used to address the physics at scales larger 
than roughly 0.03\,pc (6000\,AU). 

In the present work we use IR K-band observations of vibrationally
excited H$_2$ in the Orion Molecular Cloud (OMC1) to make a first
comparison between 
observations and hydrodynamical simulations at
the scales where individual stars are forming.
In the region observed the H$_2$ emission is optically thin in
  the sense that it is not 
self-absorbed. There will be some obscuration by dust
\citep{rosenthal}, whose spatially differential nature is not known and
is ignored here.
 The observations cover
scales from 70\,AU to 3$\cdot10^4$AU. 
OMC1 is the archetypal massive star-forming region and
the best studied region in the sky.
OMC1 is highly active with widespread on-going star
formation, exemplified by the presence of protostars, outflows and
larger scale flows \citep[see][ and references therein]{nissen2005}. 
In a previous paper
\citep{gustafsson2005}, using the same observational data as in the
present work, we quantified the nature of turbulence in OMC1 by
calculating  
size-linewidth relations, probability distribution functions and 
structure functions. It was shown that the turbulence at the small
scales covered in these data generally follows the trends observed
in CO data at larger scales. 
However, analysis also showed clear deviations from the fractal
scaling observed 
at larger scales. These deviations could be ascribed to the presence
of star formation and associated structures such as circumstellar
disks. Here we use the 
structure functions of the radial velocities and the scaling
of the structure function 
exponents from our observational data to compare with a numerical
simulation of supersonic, 
compressible, hydrodynamic turbulence. 
The scaling of structure functions has earlier been analysed
  in \cite{padoan2003} where the column densities of $^{13}$CO were used.

The structure function of order $p$ of the velocity vector $\vec{u}$
is defined here as   
\begin{equation}
S_p(r)=\langle \left|[\vec{u}(\vec{x})-\vec{u}(\vec{x}-\vec{r})]
\cdot\vec{e}\right|^p \rangle \propto r^{\zeta_p}
\label{eq:struc}
\end{equation}
where $\vec{e}$ is a unit vector parallel (longitudinal structure function) or 
perpendicular (transversal structure function) to the vector $\vec{r}$,
and $r=|\vec{r}|$. The
average is taken over all spatial positions $x$. 
The modulus sign in our definition (\ref{eq:struc}) is adopted to
improve the statistics for odd moments.
In our case the data consist of projected radial velocities. We
measure differences in radial velocity across the plane of the sky and
we are therefore dealing 
with transversal structure functions.  
The structure
functions of fully developed turbulent fields are known to follow
power laws in the 
inertial range, $\eta 
\ll r \ll L$, where $\eta$ is the dissipation scale and $L$ is the
integral scale. The set of scaling exponents, $\zeta_p$ in Eq.~(\ref{eq:struc}), can therefore be
determined \citep{frisch1995}.
The scaling exponents are expected to be characteristic of the
turbulence involved and universal for all scales and Reynolds
numbers. 
The transversal and longitudinal 
structure functions are anticipated to have the same scaling in the
infinite Reynolds number limit. This may not however be the case at
moderate Reynolds numbers, where it has been found that
$\zeta_{p,{\rm long}} > \zeta_{p,{\rm trans}}$
for $p >$~3 in incompressible hydrodynamical experiments and simulations \citep[][ and references therein]{kerr2001}.

\cite{kolmogorov1941} found from the energy conservation
law in incompressible, isotropic and homogeneous turbulence that $\zeta_3$=1.
However, \cite{dubrulle1994}
suggested that ratios of scaling exponents, say
$\zeta_p/\zeta_3$, are inherently universal, while the individual scaling
exponents may not be universal themselves. 
In this connection \cite{frick1995} showed in the context of cascade
models that one may have $\zeta_3 \neq 1$ and yet
recover scaling laws for the structure functions in good agreement
with the She-Leveque model (see below) for the ratio $\zeta_p/\zeta_3$.

\cite{she1994} described the scaling of velocity structure functions
in incompressible turbulence by:
\begin{equation}
\zeta_p/\zeta_3={p\over9} + 2\left[1-\left(\frac{2}{3}\right)^{p/3}\right],
\label{eq:she-leveque}
\end{equation}
which is confirmed by simulations of nearly incompressible turbulence
\citep{padoan2004,haugen2004a}
and by experiments \citep{anselmet1984,benzi1993}.
For supersonic turbulence \cite{boldyrev2002a} obtained, as an
extension to the She-Leveque model, the scaling:
\begin{equation}
\zeta_p/\zeta_3={p\over9} + 1-\left(\frac{1}{3}\right)^{p/3},
\label{eq:boldyrev}
\end{equation}
which is confirmed by observations in the Perseus and Taurus molecular
clouds \citep{padoan2003} and
simulations \citep{boldyrev2002b,boldyrev2002c,padoan2004}.
This type of scaling was originally proposed by \cite{politano1995}
for magnetohydrodynamic turbulence, where the dissipative structures
are thought to be two-dimensional current sheets.

In Sect.~\ref{sec:obs} we describe the observations and calculate the
structure functions from the observed velocities. We then use the
method of extended self-similarity (ESS) of \cite{benzi1993} to find the
structure function exponents and show that the scaling of these does
not represent any known theoretical scaling as represented by
Eqs.~(\ref{eq:she-leveque}) and (\ref{eq:boldyrev}). In Sect.~\ref{sec:sim} we
describe the simulation. In Sect.~\ref{sec:sim_3d} we calculate the
longitudinal and transversal structure functions of the 3D simulation
and show that the scaling of the structure function exponents is
similar to that of \cite{boldyrev2002a} in contrast to the
observations. In Sect.~\ref{sec:subset} we 
choose
subsets of the simulations which select the shocked gas seen in
the observations, project the 
data onto 2D maps and calculate the structure functions. We show that if
only strong shocks are included in the subset the scaling of the
exponents is now similar to the scaling found in the observations. In
Sect.~\ref{sec:conclusion} we discuss our results.

\section{Observations}
\label{sec:obs}

\subsection{Data}

\begin{figure*}
\centering
\includegraphics[width=13cm]{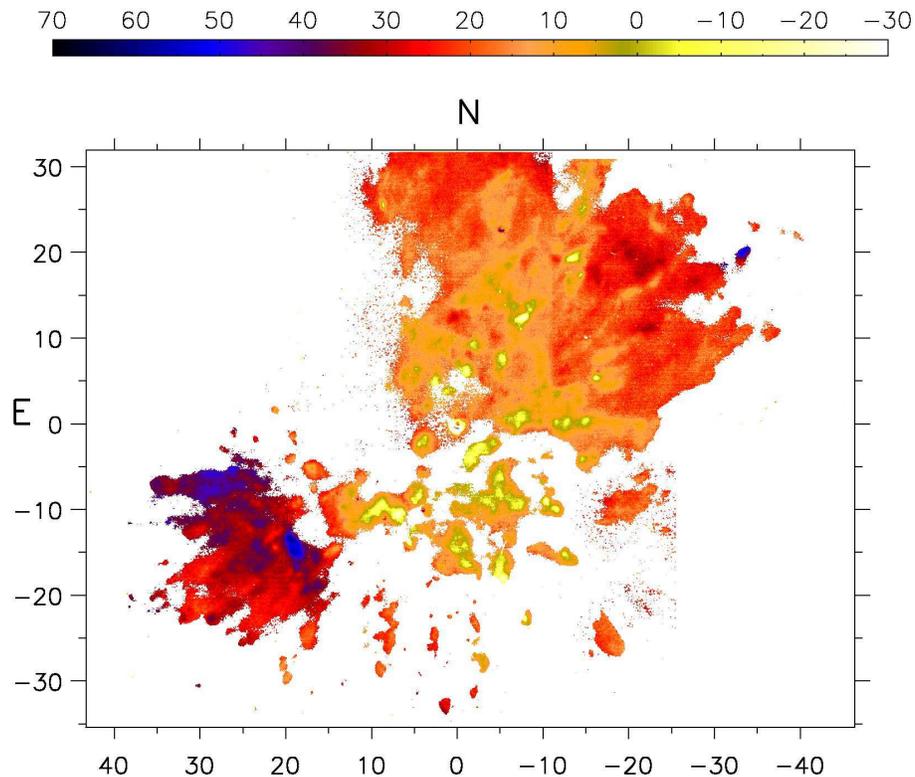}
\caption{\small The velocity field for the full observed region of
  OMC1. (0,0) is the position of the Becklin-Neugebauer object (BN)
  05$^{\mathrm{h}}$35$^{\mathrm{m}}$14\fs12, $-05\degr22'22\farcs9$
  (J2000). Axes are labelled in arcseconds and colours represent radial
  velocities in $\kms$.
  For the white regions no data are available.}
\label{fig:vel_obs}
\end{figure*}

We use the radial velocity map of the BN/KL region of the Orion Molecular Cloud
(OMC1) of \cite{gustafsson2003,gustafsson2005,nissen2005}. 
The data contain both brightness and velocity information and were 
obtained at the CFHT with a Fabry-Perot interferometer in 
conjunction with adaptive optics using the so-called GriF instrument
\citep{clenet}. Observations were performed
in the NIR K-band by scanning the $v$=1-0 S(1) H$_2$ emission line at
2.121$\mu$m. The field of view is 36\arcsec $\times$ 36\arcsec and the
pixel scale is 0\farcs035 where 1\arcsec $= 460\,$AU. The 
dataset consists of four 
spatial and velocity resolved images, which are amalgamated into one
field of 89\arcsec$\times$67\arcsec or 0.2$\times$0.15\,pc for a
distance of Orion of 460\,pc \citep{bally}. The field is centered
approximately on the Becklin-Neugebauer (BN) object
(05$^{\mathrm{h}}$35$^{\mathrm{m}}$14\fs1, $-05\degr22\arcmin 
22\farcs9$), see Fig.~\ref{fig:vel_obs}. The spatial resolution is
0\farcs15 (70\,AU). The radial velocity at each spatial
position was determined by the peak position in a lorentzian fit to
the velocity profile provided by the Fabry-Perot. Relative velocities
are determined with an accuracy of between $1\kms$ (3$\sigma$) in the brightest regions and 
$8$--$9\kms$ in the weakest regions considered here.
Systematic errors due to mechanical instabilities in the Fabry-Perot
may occur in establishing velocity differences 
between distant regions. As discussed in \cite{gustafsson2005}, tests
have been performed to show that such systematic errors are not
present in the data to any significant extent.

The emission of vibrationally
excited H$_2$ observed here does not trace the bulk of the gas. Rather it
traces hot, dense 
gas, where excitation occurs largely through 
shock excitation. 
Detailed models \citep{storzer} show that the maximum brightness in
the H$_2$ $v$=1--0 s(1) line from
fluorescence in this region of OMC1 does not exceed 10--15\% of the
total brightness observed \citep{kristensen2003}. Thus photon
excitation is a minor contributor to the total brightness.

\subsection{Results from observations}

Since observational data provide only radial velocities in the plane of the
sky in a 2D projection
of the gas motions, 
 we obtain only the
transversal structure functions, as described earlier.
Furthermore the accuracy of the velocity data in any
pixel depends on the brightness in that pixel. 
\cite{gustafsson2005} showed that more robust structure functions are obtained when the velocity differences are weighted by
the brightness. On this basis we use a modified
definition of the structure functions:  
\begin{equation}
S_p(r)=\langle B(\vec{x})B(\vec{x}-\vec{r})\,
|v(\vec{x})-v(\vec{x}-\vec{r})|^p \rangle.
\label{eq:weigh}
\end{equation}
Here $v$ is the line of sight velocity and the average is extended over all
spatial positions $\vec{x}$ and all lags $\vec{r}$ where $r=|\vec{r}|$.
$B(\vec{x})$ is the brightness at position $\vec{x}$. We thus weight each
velocity difference by the product of the brightness of the two
spatial positions involved, thereby giving more weight to the
brightest regions which exhibit the highest accuracy in the radial
velocity.

\begin{figure}
\includegraphics[width=7.8cm]{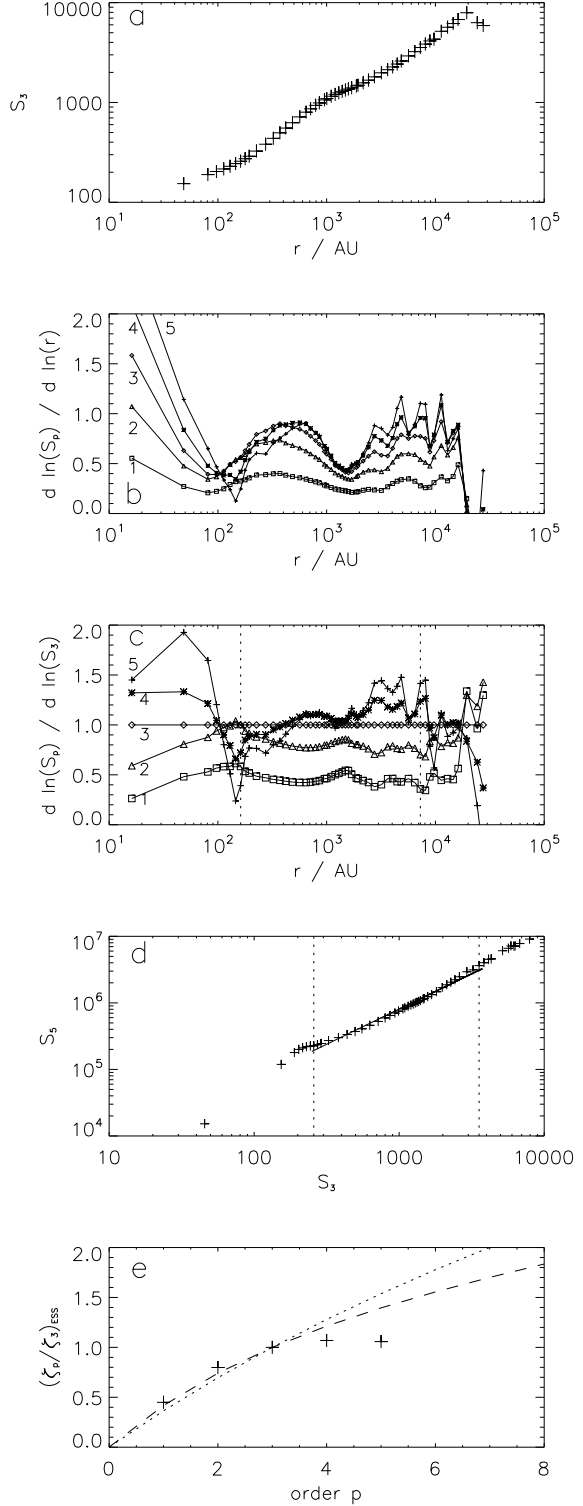}
\caption{\small a) Third order structure function of the velocities in
  OMC1. b) Logarithmic derivatives of $S_p(r)$ for
$p=1$--$5$ c) Ratios of the differential slopes of $S_P(r)$ to the slope of the third
  order structure function for $p=1$--$5$. The vertical dotted lines mark the
  interval in which the scaling exponents have been fitted. d) $S_5(r)$
  vs.\ $S_3(r)$. The dotted lines mark the range of the fit and the
  solid line is the best fit within that range yielding the
  logarithmic slope, $(\zeta_p/\zeta_3)_{\rm ESS}= 1.06$. e) The 
ESS scaling exponents  (+) OMC1, (dotted line)
  She-Leveque scaling, (dashed line) Boldyrev scaling.}
\label{fig:scal_obs}
\end{figure}

The third order structure function of OMC1, $S_3(r)$, is displayed in
Fig.~\ref{fig:scal_obs}a. It is not well represented by a single power
law showing a clear deviation around 10$^3$AU. This is also evident
from the large variations in the local logarithmic derivatives
of $S_p(r)$, shown in Fig.~\ref{fig:scal_obs}b for $p=1$--$5$,
where the derivatives are evaluated numerically using a three-point formula.
However, \cite{benzi1993} discovered that the structure functions can be
represented as functions of, say, the third order structure function, namely
\begin{equation}
S_p(r) \propto S_3(r)^{(\zeta_p/\zeta_3)_{\rm ESS}}.
\label{eq:ess}
\end{equation}
This is now known as extended self-similarity (ESS).
Even if the
structure functions of Eq.~(\ref{eq:struc}) are not power laws over any given range, the
functions represented by Eq.~(\ref{eq:ess}) nevertheless exhibit good power
law behaviour. The scaling in Eq.~(\ref{eq:ess}) is generally found to
extend over a much larger range than
for the structure functions of Eq.~(\ref{eq:struc}).
As emphasized before, self-similarity, as expressed by 
Eq.~(\ref{eq:ess}), is believed to be more fundamental than the
self-similar scaling 
with respect to $r$ \citep{benzi1993}. 
In Fig.~\ref{fig:scal_obs}c we have
plotted the ratio of the logarithmic slopes of $S_p$ and $S_3$,
$d\ln S_p(r)/d\ln S_3(r)$, for $p=1$--$5$. If a range 
in which good power law scaling is
present is encountered in the various structure functions, the ratio of logarithmic
slopes should display plateaus in that range
at values of $(\zeta_p/\zeta_3)_{\rm ESS}$. From Fig.~\ref{fig:scal_obs}c we
find that the structure functions for $p=1$--$5$ exhibit a reasonably good
scaling range from $r=160\,$AU to $r=7000\,$AU. This range is marked by the dotted
vertical lines in Fig.~\ref{fig:scal_obs}c. The scaling exponents
are found by fits to Eq.~(\ref{eq:ess}) in this range. As an
  example we show in Fig.~\ref{fig:scal_obs}d the extended
  self-similarity plot of $S_5(r)$ vs.\ $S_3(r)$ together with the
  best fit yielding the slope $(\zeta_p/\zeta_3)_{\rm ESS} =
  1.06$. The dotted lines mark the range of the fit. It is however clear
from Fig.~\ref{fig:scal_obs}c that the scaling gets poorer
when the order $p$ is increased. At 
$p=5$ the plateau is rather poorly defined (see also Fig.~\ref{fig:scal_obs}d) and therefore we cannot determine a
scaling at higher orders than $p=5$. In Fig.~\ref{fig:scal_obs}e we show
the scaling exponents $(\zeta_p/\zeta_3)_{\rm ESS}$ vs.\ $p$ (+
signs) compared to the
values predicted by the She-Leveque model of incompressible
turbulence, Eq.~(\ref{eq:she-leveque})
(dotted line) and the Boldyrev model of supersonic
turbulence, Eq.~(\ref{eq:boldyrev}) (dashed line). The scaling exponents derived from
the velocity in OMC1 clearly deviate from both the She-Leveque
and the Boldyrev scaling at $p\geq4$. The OMC1 scaling exponents show
signs of becoming constant at $(\zeta_p/\zeta_3)_{\rm ESS}\sim 1$ or even
slightly decreasing for $p>4$, in contrast to the theoretical
scalings, which are monotonically increasing. 

This result for velocities of hot, shocked gas in OMC1 at scales
$70\,$AU -- 3$\cdot 10^4$AU (3.4$\cdot
10^{-4}$pc to $0.15\,$\,pc) differs from the findings of 
\cite{padoan2003}. They found that the density fields in the Perseus and
Taurus molecular clouds as observed in CO follow Boldyrev scaling at
scales larger than $0.08\,$pc.

Below we will show that the unusual scaling found here can be
reproduced by 
numerical simulations of supersonic turbulence when only subsets of
the simulations representing the shocked regions are considered.

\section{Simulations}
\label{sec:sim}
In order to understand some of the peculiar scalings found in the
observations we now consider data of supersonic isothermal
compressible turbulence
simulations.
Such simulations have been performed by a number of different groups
\citep{passot1987,vazquez1995,padoan1998,klessen2000,vazquez2003,cho2003,kritsuk2004}.
Here we consider simulations that are most closely related to those
of \cite{haugen2004b}, except that magnetic fields are neglected here.
The governing equations are
\begin{equation}
{\partial\vec{u}\over\partial t}+\vec{u}\cdot\vec\nabla\vec{u}
=-c_{\rm s}^2\vec\nabla\ln\rho+\vec f
+{1\over\rho}\vec\nabla\cdot\vec\tau,
\end{equation}
\begin{equation}
{\partial\ln\rho\over\partial t}+\vec{u}\cdot\vec\nabla\ln\rho
=-\vec\nabla\cdot\vec{u},
\end{equation}
where
$\tau_{ij}=2\rho\nu{\sf S}_{ij}+\rho\mu\delta_{ij}\vec\nabla\cdot\vec{u}$
is the stress tensor and
${\sf S}_{ij}={1\over2}(u_{i,j}+u_{j,i})-{1\over3}\delta_{ij}u_{k,k}$
is the rate of strain matrix and commas denote partial differentiation.
Following \cite{nordlund1995}, we assume $\mu$ to be
proportional to the smoothed and broadened positive part of the negative
divergence of the velocity, i.e.\ 
\begin{equation}
\mu=c_{\rm shock}\left<\max_5[(-\vec\nabla\vec{u})_+]\right>,
\label{eq:c-shock}
\end{equation}
where $c_{\rm shock}$ is the artificial viscosity parameter and the
5 underneath the $\max$ operator indicates that the
maximum is taken over 5 by 5 by 5 mesh widths (or ~"pixels").
This is also the technique used by \cite{padoan2002} and \cite{haugen2004b}.
The function $\vec{f}$ denotes a random forcing function that consists of
plane waves, normalized by a dimensionless amplitude factor $f_0$ that will be
varied in the different simulations discussed below (see Appendix~A).

The equations are solved on a periodic mesh of size $L^3$, where
$L=2\pi/k_1$ is the length of the side of the box and $k_1$ is the smallest
wave number in the domain.
We use the {\sc Pencil Code}, which is a high-order
finite-difference code (sixth order in space and third order in time)
for solving the compressible hydrodynamic equations.\footnote{
\url{http://www.nordita.dk/software/pencil-code}}

We consider runs with different forcing amplitudes, $f_0$,
leading to different root mean square Mach numbers,
$\mbox{Ma}_{\rm rms}=u_{\rm rms}/c_{\rm s}$ where $c_{\rm s}$ is the
speed of sound, see Table~\ref{tab:sims}.
The high resolution runs, in rows 2 and 3 of Table~\ref{tab:sims}, have been evolved for about 40 sound travel times,
$\tau_{\rm sound}=(c_s k_1)^{-1}$, while the low resolution run
has been conducted for about $90\,\tau_{\rm sound}$.
The sound travel time can be associated with the turnover time by
noting that $\tau_{\rm turn}=(u_{\rm rms}k_1)^{-1}=\tau_{\rm sound}/\mbox{Ma}_{\rm rms}$. 

\begin{table}
\caption{\small Parameters of the numerical simulations. Resolution,
shock viscosity, forcing amplitude, Mach number, run time $\Delta t_{\rm run}$ in terms of
turnover times $\tau_{\rm turn}$.}
\label{tab:sims}
\begin{tabular}{cccccc}
\hline
\hline
run&resolution&$c_{\rm shock}$&$f_0$&
${\rm Ma}_{\rm rms}$&$\Delta t_{\rm run}/\tau_{\rm turn}$\\
\hline
1&$256^3$&2& 2& 3   &270\\ 
2&$512^3$&3&10&7--9 &320\\ 
3&$512^3$&2&10&8--10&360\\ 
\hline
\end{tabular}
\end{table}

\subsection{Results from full 3D simulation}
\label{sec:sim_3d}

First, we show that the structure functions of the full 3D
simulation follow the theoretical scaling of \cite{boldyrev2002a}.
For a snapshot of Run~1 at $t=70\,\tau_{\rm sound}$
(corresponding to $t=210\,\tau_{\rm turn}$) 
we have calculated the longitudinal and
transversal structure functions of the 3D simulation using 
\begin{eqnarray}
S_{p,\rm long}(l)&=&\langle |u_x(x+l,y,z)-u_x(x,y,z)|^p \rangle \label{eq:long},
\\ 
S_{p,\rm trans}(l)&=&\langle |u_y(x+l,y,z)-u_y(x,y,z)|^p \rangle\nonumber\\
&+&\langle |u_z(x+l,y,z)-u_z(x,y,z)|^p \rangle \label{eq:trans},
\end{eqnarray}
as in Eq.~(\ref{eq:struc}).

\begin{figure}
\centering
\includegraphics[width=8cm]{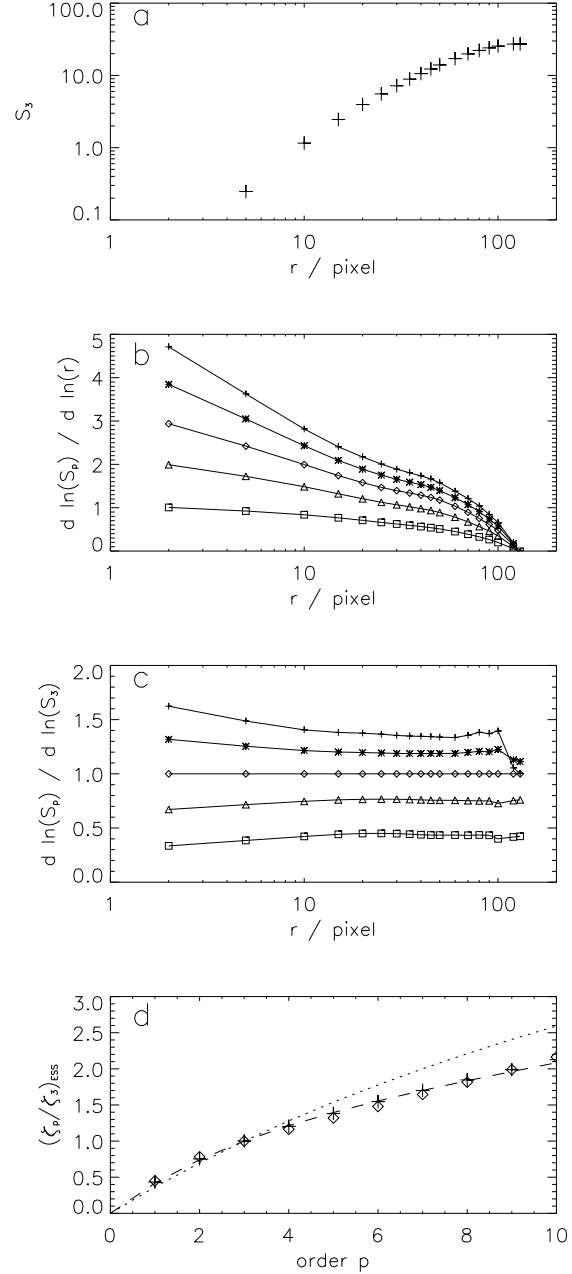}
\caption{\small a) The third order transversal structure function of
  Run~1. b) logarithmic derivatives of $S_p(r)$ for
$p=1$--$5$ (in ascending order). c) Ratios of the differential slopes of the
  transversal structure functions of order 1-5 to the
  slope of the third order structure function. d) The ESS scaling exponents
  of the transversal structure function
  (+) and the 
  longitudinal structure function (diamonds) compared to the
  She-Leveque scaling (dotted line) and the Boldyrev scaling
  (dashed line).}
\label{fig:scal_3d}
\end{figure}

In Fig.~\ref{fig:scal_3d}a the third order transversal structure
function is shown. The logarithmic derivatives of $S_{p,\rm
  trans}(l)$ (Fig.~\ref{fig:scal_3d}b) show no range of scales where
plateaus (good power law scaling) are present. Note, however,
  that in these simulations $\zeta_2$ is closer to unity than $\zeta_3$.
In Fig.~\ref{fig:scal_3d}c we have plotted the
logarithmic slope of $S_{p,\rm trans}(l)$ vs.\ $S_{3,\rm trans}(l)$ for
$p=1$--$5$, i.e.\
again using the method of extended self-similarity (ESS). A range of good scaling is
now seen to be present over most of the dynamical range from 10--80
mesh widths. The longitudinal structure functions are nearly identical to
the transversal and are not shown here.
The scaling exponents found from fits to the structure functions in
the interval of 10--80 are plotted in Fig.~\ref{fig:scal_3d}d for both
the transversal structure functions (+) and the longitudinal
structure functions (diamonds). Both the transversal and the
longitudinal structure functions follow the velocity scaling for
supersonic turbulence of \cite{boldyrev2002a}.

\subsection{Scaling of subsets of the simulations}
\label{sec:subset}

\begin{figure*}[]
\centering
\includegraphics[width=17.5cm]{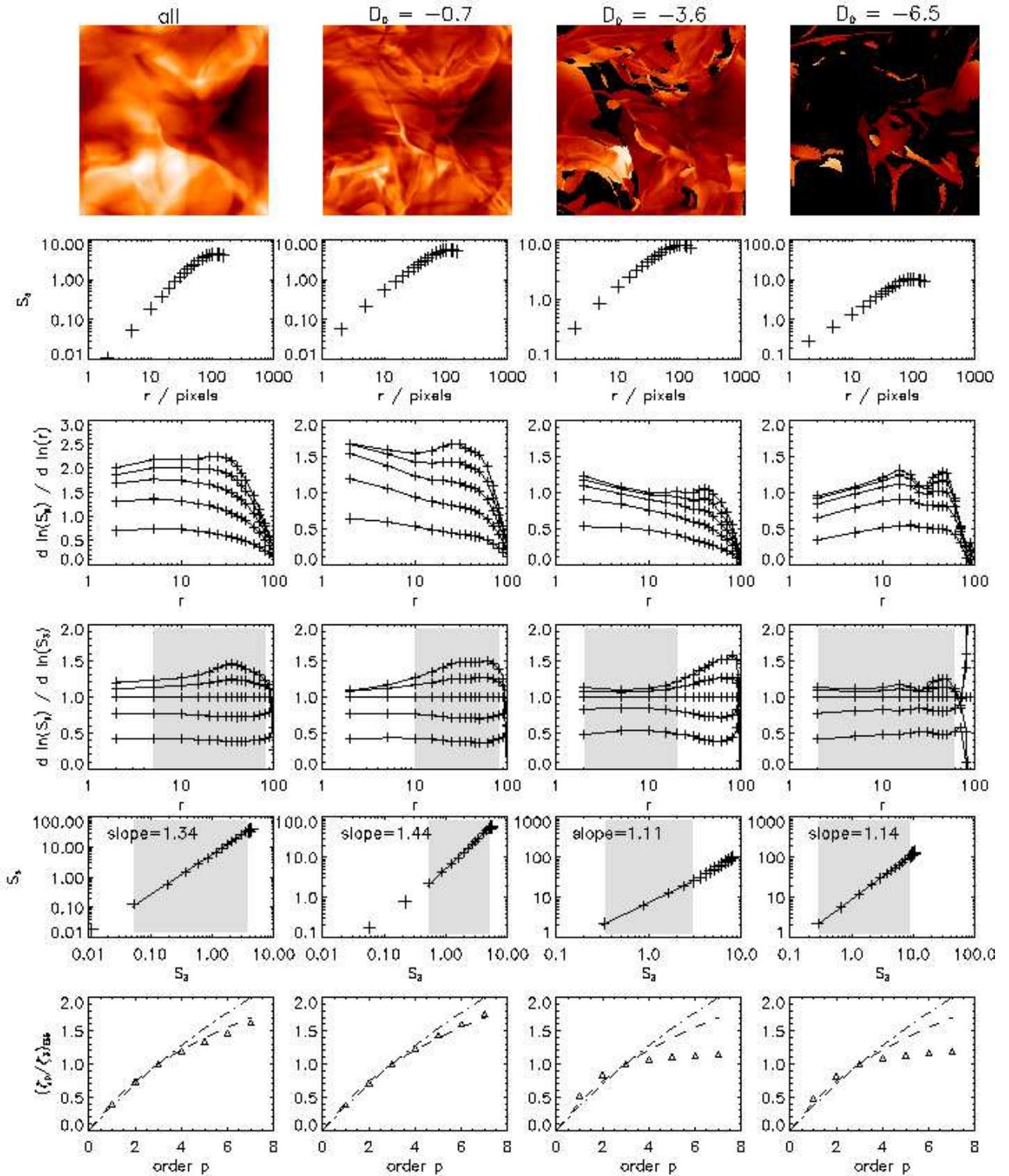}
\caption{\small Run~1, $t=50\,\tau_{\rm sound}$. Results for projected
  maps including all points (1st column) and
for subsets with $D_0=-0.7$, $-3.6$, and
$-6.5$ (as marked), using $w=0.7$. Top
row: radial velocity maps projected in the $z$-direction. Second row: third
order structure functions averaged over 50 projection angles (see
text). Third row: logarithmic derivatives of $S_p(r)$ for
$p=1$--$5$ (in ascending order).
Fourth row: ratios of differential slopes to $\zeta_3$ for order
$p=1$--$5$.
The grey shades indicate the ranges over which average values of
$(\zeta_p/\zeta_3)_{\rm ESS}$ are determined.
Fifth row: $S_5(r)$ vs.\ $S_3(r)$.
The ranges of grey shades correspond to those in the fourth row.
Solid lines are best fits within the
indicated ranges yielding the logarithmic slope, $(\zeta_p/\zeta_3)_{\rm
  ESS}$.
Bottom row: the radial velocity scaling compared to the
She-Leveque scaling (dotted line) and the Boldyrev scaling
(dashed line).} 
\label{fig:sim_sub1}
\end{figure*}

We now study structure functions of subsets of the simulations
and select subsets that resemble best the physical properties of the
observational data, that is being composed of preferentially shocked gas. 
Similar work has already been carried out by \cite{kritsuk2004}.
First, in order to compare the simulations to the observations, we need to
project the simulated 3D velocity components onto a 2D
map of only radial velocity. The radial velocity in each spatial
position is found by averaging the density
weighed $z$-component (say) of the velocity over the $z$-range. That is, 
\begin{equation}
\overline{u}_z(x,y)=\left.\int_z\rho u_z\,{\rm d}z\right/\int_z \rho\,{\rm d}z.
\label{eq:avr}
\end{equation}  
We have checked that this expression yields the same values of
  velocities as the method adopted in the reduction of the observational
  data obtained with GriF. In the
  observations, the true H$_2$ line profile is convolved with the very
much broader instrumental
  Lorentzian profile of the Fabry-Perot interferometer and the radial
  velocity is found from a Lorentzian fit. The same procedure has been
  used on simulated velocity profiles through convolution and
  fitting and it has been found in numerous tests that the velocities
derived are essentially the same as the centroid
  velocities obtained via Eq.~(\ref{eq:avr}).
In Fig.~\ref{fig:sim_sub1} (top left, marked "all") the resulting 2D map is
shown for Run~1 at $t=50\,\tau_{\rm sound}$,
corresponding to $t=150\,\tau_{\rm turn}$.

If the turbulence is homogeneous and isotropic the
projected map should be independent of the projection
angle. However, since the simulations have limited spatial extent, they
turn out to show residual anisotropy, in the sense that independence of
projection angle is not assured. Thus the projection map and
subsequently the structure functions could
depend on the projection angle. To minimize such effects we have
calculated the structure functions for a number of random projection
angles. An average of 50 angles was taken for Run~1. The higher
resolution of Runs~2 and 3 should alleviate the problem of projection
angle and averages over only 3 angles were taken in these cases. In the following all structure functions of 
projected maps refer to an average of structure functions. There
could also be projection effects in the observations, but we
have no choice but to ignore these. 

The average third order structure 
function [Eq.~(\ref{eq:weigh}) with $p=3$], the logarithmic
derivatives of $S_p(r)$ for $p=1$--$5$, the  
ratio of the logarithmic slopes of $S_p(r)$ and $S_3(r)$, the
  extended self-similarity plot of $S_5(r)$ vs.\ $S_3(r)$, and
the velocity scaling exponents are also shown in
Fig.~\ref{fig:sim_sub1}, passing down the left column, for the
projected simulation of Run~1.
In calculating the structure functions of the
simulations we use $B=1$, that is, no brightness weighting in
Eq.~(\ref{eq:weigh}) since the velocities in 
the simulations are free of `observational' errors.
It is found that the scaling of the structure functions of the projected
radial velocity follows that of Boldyrev, as did the transversal and
longitudinal structure functions of the full 3D simulation
(Fig.~\ref{fig:scal_3d}).  

The problem is to identify the subset of structures in the simulations
which corresponds to the structures represented in our observations. 
As mentioned above we observe a subset of the gas in OMC1 consisting
very largely of shocked gas. 
In order to make comparison between
observations and simulations it
is therefore necessary to extract regions  in the
simulations where shocks occur. Shocks are generated where fast
material attempts to 
overtake slower moving material and material shows rapid
deceleration, that is, where a negative velocity 
gradient is present. In simulations, shocked regions can thus be
distinguished as 
regions with suitably strong negative divergence ($\vec{\nabla}\cdot\vec{u}<0$),
that is, convergence.    

\begin{figure}[]
\centering
\includegraphics[width=8cm]{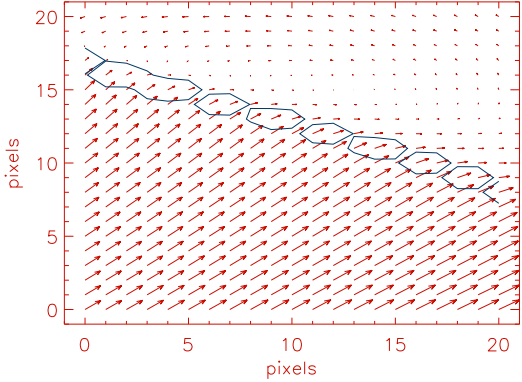}
\caption{\small Vectorplot of ($u_x,u_y$) in a section of an
  $xy$-plane in Run~1. Contours outline shocked regions where $s>0$ for the
  cut-off value $D_0 =-6.5$, $w=0.7$, that is, regions where $D < -5.8$.}
\label{fig:vector}
\end{figure}

Thus, in order to 
compare with the observations, we choose different subsets of the simulations
consisting of regions with shocks stronger than a certain degree.
\cite{kritsuk2004} accomplished this by selecting regions where the
density exceeds a certain threshold value.
Here, on the other hand, we consider only regions that
have stronger convergence than a given cut-off value, $D_0<$0.
This is achieved by defining
\begin{equation}
D=\vec\nabla\cdot\vec{u}/\langle(\vec\nabla\cdot\vec{u})^2\rangle^{1/2}
\end{equation}
as the relative velocity divergence, and the selection function $s$:
\begin{eqnarray}
s(x,y,z)=\left \{ \begin{array}{lllll}
0 && \xi&\le-1 &\\
\textstyle{1\over2}+\textstyle{1\over4}\xi(3-\xi^2)
 & \quad-1<&\xi&<1 & \\
1 && \xi&\ge1, &
\end{array} \right.
\label{eq:selec}
\end{eqnarray}
where $\xi=(D_0-D)/w$, and
$w$ is the width of the selection function. All points with
$D > D_0+w$, which we seek to exclude, have
$s=0$. The selection function is chosen for its smooth variation
between 0 and 1. We have used other somewhat different forms of the selection function
and found that the results do not significantly depend on the
specific form. Figure~\ref{fig:vector} shows an example 
of a local velocity field where shocks are present. The figure shows
velocity vectors ($u_x,u_y$) in an $xy$-plane and contours of shocked regions where $s>0$ for
$D_0=-6.5$ and $w=0.7$ in Run~1.

\begin{figure*}
\centering
\includegraphics[width=0.8\textwidth]{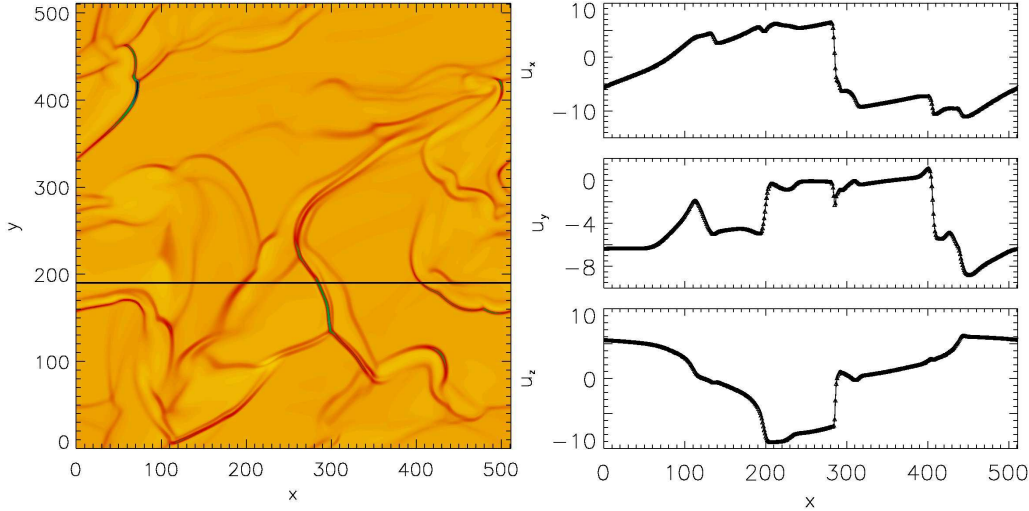}
\caption{\small Left panel: $xy$-slice of $\vec\nabla\cdot\vec{u}$ through
the box of Run~3 at some arbitrarily chosen value of $z$.
The horizontal line indicates the location along which the
profiles of $u_x$, $u_y$, and $u_z$ (in $\kms$) are shown in the three panels on
the right hand side.
A strong shock is seen at $x=290$.}
\label{fig:shock}
\end{figure*}

The shock structure in an $xy$-slice of Run~3 is shown in
Fig.~\ref{fig:shock}, where regions with large negative values of
$\vec\nabla\cdot\vec{u}$ (darker regions) represent shocks. The profiles
of  $u_x$, $u_y$, and $u_z$ are shown along a horizontal line on which
the presence of a shock for example at $x=290$ is evident through
large velocity 
changes in $u_x$, $u_y$, and $u_z$ over a range of only $\sim 5$
mesh widths. The higher Mach number 
of these simulations leads to larger velocity differences and stronger
shocks compared to the simulations of Run~1.\footnote{
Movies of the time evolution of the simulations can be found at:
\url{http://www.nordita.dk/~brandenb/movies/shockdiss/}}

The radial velocity in each spatial position $(x,y)$ is now found by a
modified form of Eq.~(\ref{eq:avr}):
\begin{equation}
\overline{u}_z(x,y)=\left.\int_z s\rho u_z\,{\rm d}z\right/\int_z
  s\rho\,{\rm d}z.
\label{eq:sub_set}
\end{equation}  
Returning to Fig.~\ref{fig:sim_sub1}, for a snapshot of Run~1 at
$t=50\,\tau_{\rm sound}$ this shows maps projected in the
$z$-direction as examples of subsets with  $w=0.7$ and $D_0=-0.7$, $-3.6$, and 
$-6.5$. 
The value of $D$ in a region depends on how the velocity changes in
the vicinity of that region. Large differences
in velocity over a limited region lead to high values of $D$. Thus the
restrictions on $D$ can be associated with typical minimum values of the
velocity change that must occur in a shocked region for that region to
be included in the structure function analysis. For example the physical
interpretation of the restriction $D_0 = -3.6$ is that in
order for a
pixel to be included there must be a velocity gradient in the
immediate vicinity of that
pixel, such that $|\Delta \vec{u}| \sim 3\kms$
over $\Delta r =10$ pixels. 
An estimate of the value of the gradient in physical units can be
given by observing that the size of shocks in the simulations is
typically 5 pixels. Assuming a physical shock width of C-shocks of
50\,AU \citep{lacombe2004}, the value of the required velocity gradient
is $\sim 0.03\kms\,{\rm AU}^{-1}$.  
When $D_0 = -6.5$, typical values are  
$|\Delta \vec{u}| \sim 5\kms$ again over $\Delta r =10$ pixels. These values are estimated for Run~1
and are found to be higher in the runs with stronger forcing and
higher Mach numbers for the same value of $D$.

The map of $D_0 = -0.7$, that is, including all shocked
regions, displays sharp filamentary structure, compared with the map of
all points in the simulation, which has smoother contours (see top row
of Fig.~\ref{fig:sim_sub1}). Excluding
the weakest shocks, that is, for $D_0 = -3.6$, leads to a
more broken up appearance, and the filaments are clearly
visible. When only the strongest shocks are considered, $D_0 = -6.5$,
the radial velocity map consists mostly of sheets and filaments. 

Figure~\ref{fig:sim_sub1} also displays the third order structure functions
for the three subsets as defined by values of $D_0$, as well as the logarithmic
derivatives of $S_p(r)$, the  
ratio of the logarithmic slopes of $S_p(r)$ and $S_3(r)$, the
  extended self-similarity plot of $S_5(r)$ and $S_3(r)$, as well as
the scaling exponents of the structure functions of the radial velocity. The
structure functions are averages of 50 
projected maps as described above. The third order structure function for the full simulation and the
three subsets all display good power laws. The leveling off of the
power laws at lags around
100 mesh widths is an artifact due to the finite size of the simulation of size 256. No
bumps in the structure functions are seen. This is in marked contrast
to the structure functions of the
observations where clear bumps are present
(see the first panel of Fig.~\ref{fig:scal_obs}).

The ratios of logarithmic slopes show plateaus over about an
order of magnitude in scale,
especially in the lower order structure functions, $p=1$, $2$.
The scaling exponents
$\zeta_p/\zeta_3$ are found by fits to Eq.~(\ref{eq:ess}) in the
interval of $r$ where  $d\ln S_p(r) / d\ln S_3(r) $ shows the best
plateau.  
The plateau is found at $r=5$ to 80 mesh widths for the full simulation, at
$r=10$ to 80 mesh widths for $D_0 = -0.7$, at $r=2$ to 20 for $D_0 = -3.6$ and at
$r=2$ to 60 for $D_0 = -6.5$. The ranges used for fitting are indicated
in Fig.~\ref{fig:sim_sub1} with grey shading.
The best power law fits to $S_5(r)$
  vs.\ $S_3(r)$ in the indicated ranges are shown in
  the fifth row of  Fig.~\ref{fig:sim_sub1}. The value of the slope is
  indicated. 
The velocity
scaling is close to following that of Boldyrev when all points in the
simulation are included and when only shocked regions with $D < 0$
($D_0 = -0.7$) are
included. There is, however, a dramatic change in the scaling when the
restrictions on the strength of the shocks are made tighter. For
both $D_0 = -3.6$ and $D_0 = -6.5$ the scaling
deviates strongly from both She-Leveque and Boldyrev
scalings. 
This change in behaviour is associated with a shift in the range for
which a plateau can be seen. Especially in the last column of
Fig.~\ref{fig:sim_sub1} the scaling
is seen to extend all the way from the resolution limit (2 mesh widths) to
60 mesh widths. 
This cannot be regarded as regular inertial range scaling because the
usual dissipative subrange, always present in turbulence
simulations, cannot be distinguished.
We associate the apparent shift of the ranges with the presence of shock
structures which become more strongly pronounced as the cutoff value,
$D_0$, is moved to more negative values.
Thus we are beginning to see effects due to the use of artificial viscosity
when $D$ becomes sufficiently negative.
These artificially smoothed shocks may resemble C-type shocks
that occur in the magnetized interstellar medium and are known to be a
common feature in OMC1 \citep{gustafsson2003, nissen2005}.
We also note that the nominal dissipation scale without artificial viscosity
is just 1--2 mesh widths, which is still below the artificial dissipation scale of the
shocks of $\sim$~5 or more mesh widths.
The scaling is seen to remain nearly constant for $p\geq3$,
resembling the scaling found in the observations of OMC1.
This is qualitatively similar to the results of \cite{kritsuk2004},
who found systematically smaller scaling exponents for $p>3$ when
only high density regions are considered.
This shows
that the unusual scaling observed in OMC1 can be the effect of observing
only the hot, shocked gas, and that hydrodynamical turbulence
simulations without self-gravity or magnetic fields are able to reproduce this. 

\begin{figure*}
\centering
\includegraphics[width=17.5cm]{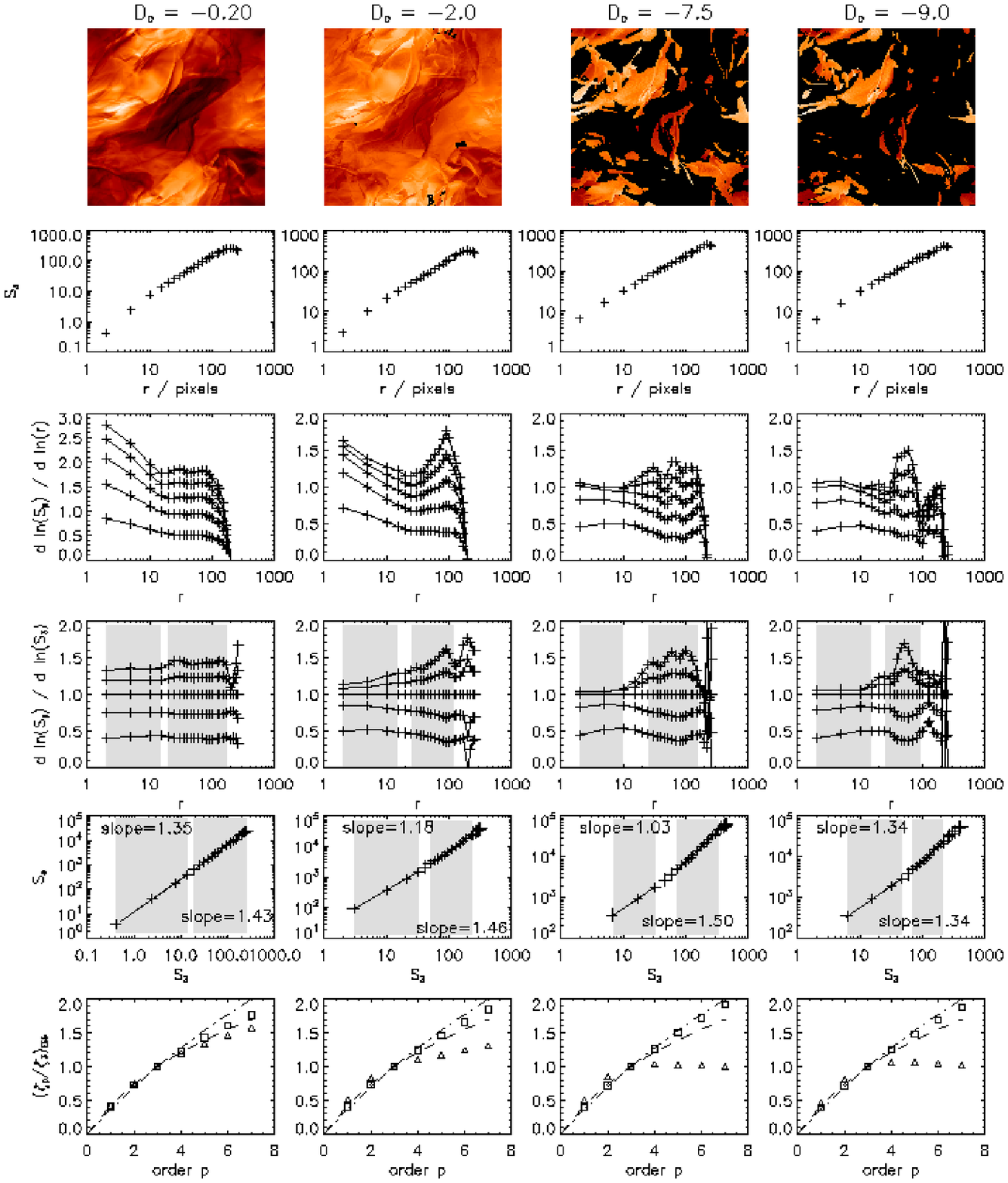}
\caption{\small  Run~3, $t=39\,\tau_{\rm sound}$. Results for projected maps 
for subsets with $w=0.20$ and $D_0=-0.20$, $-2.0$, $-7.5$ and
$-9.0$. Top
row: radial velocity maps projected in the $z$-direction. Second row: third
order structure functions averaged over 3 projection angles (see
text). Third row: logarithmic derivatives of $S_p(r)$ for
$p=1$--$5$ (in ascending order).
Fourth row: ratios of differential slopes to $\zeta_3$ for order
$p=1$--$5$.
The grey shades indicate the ranges over which average values of
$(\zeta_p/\zeta_3)_{\rm ESS}$ are determined.
Fifth row: $S_5(r)$ vs.\ $S_3(r)$.
The ranges of grey shades correspond to those in the fourth row.
Solid lines are best fits within the
indicated ranges yielding the logarithmic slope, $(\zeta_p/\zeta_3)_{\rm
  ESS}$. 
Bottom row: the radial velocity scaling compared to the
She-Leveque scaling (dotted line) and the Boldyrev scaling
(dashed line). $\triangle$: large scales, $\Box$: small scales.}
\label{fig:sim_512h}
\end{figure*}

Figure~\ref{fig:sim_512h} shows similar results to Fig.~\ref{fig:sim_sub1} but
for a snapshot of Run~3 at $t=39\,\tau_{\rm sound}$.
The subsets
corresponding to $D_0=-0.20$, $-2.0$, $-7.5$, and $-9.0$
show the same trends as seen in Fig.~\ref{fig:sim_sub1}. The tighter
the restrictions on the strength of the shocks, the sharper and more
broken up the 
filamentary structures become. The radial velocity scaling flattens
strongly when
$D_0 = -7.5$ and $-9.0$ (bottom row in Fig.~\ref{fig:sim_512h}).

The grey-shaded areas in the fourth row of Fig.~\ref{fig:sim_512h} indicate
 two different regions where scaling can tentatively be noted, in
 contrast to the single regions identified in
 Fig.~\ref{fig:sim_sub1}. 
As noted earlier, the dissipation scale for low velocity
  gradients is essentially given by 
1--2 mesh widths. This is equivalent to the case in which no artificial
viscosity is introduced; see Eq.~(\ref{eq:c-shock}). The dissipation
scale rises to 5 or more
mesh widths in locations where large velocity gradients are encountered. The two
different scales identified in Fig.~\ref{fig:sim_512h} (rows 4 and 5)
have dimensions 
of 2 to $>$10 mesh widths and 30 to $>$100 meshwidths. 
The smaller of these ranges covers that associated with artificial
viscosity. The scaling at the lower range will thus be affected by the
presence of shocks. This may take place through the locally enhanced
viscosity embodied in  Eq.~(\ref{eq:c-shock}).

We now consider the scaling associated with the two different regions
separately. In Fig.~\ref{fig:sim_512h}, open triangles refer to the
larger regions and open squares to the smaller regions.
For the data in the left column for $D_0 =-0.20$, where all regions with a
positive value of the convergence are included, that is, all shocks, it is seen that both
scaling regions (small scales, 2--15 mesh widths, and large scales, 20--170
mesh widths) show scaling behaviour roughly compatible with Boldyrev scaling.
When we introduce more restrictive thresholds for
$D_0$ different behaviour is found. The large scales then begin to
follow more closely the standard She-Leveque scaling. At the same time,
the small scales, associated with shocks, show the levelling off discussed
in connection with Fig.~\ref{fig:sim_sub1} for strongly shocked
regions, and as seen in the 
observations, Fig.~\ref{fig:scal_obs}e. We therefore are able to
explain the unexpected 
scaling found in the observations through an inherent selection of shocked
regions.

We note that with the greater energy dissipation associated with
Run~2, slopes for 
small and large scales differ as for Run~3 but less markedly.

\begin{figure}
\centering
\includegraphics[width=\columnwidth]{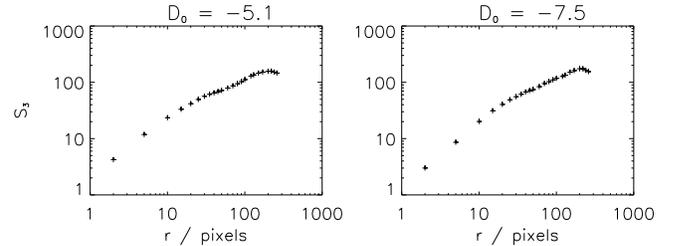}
\caption{\small Run~2, $t=37\,\tau_{\rm sound}$. Third order structure
functions of projected maps with $w=0.2$ and $D_0=-5.1$ and $D_0=-7.5$.}
\label{fig:sim_512g}
\end{figure}

\subsection{Non-power law behaviour in structure functions}
The structure functions of Run~3 (second row in
Fig.~\ref{fig:sim_512h}) are all well represented by power laws in
contrast to the behaviour found in the structure
functions of OMC1 (Fig.~\ref{fig:scal_obs}). In one snapshot of Run~2,
however, similar bumps to those
in the observations are detected; see Fig.~\ref{fig:sim_512g}. The
deviations from power 
laws of the structure functions in Run~2 are found in subsets with high
values of the threshold, $D_0$, in the snapshot at $t=37\,\tau_{\rm
  sound}$. Examples
are seen in Fig.~\ref{fig:sim_512g} for $D_0 = -5.1$ and $D_0 = -7.5$ where
it is clear that the third order structure
functions displays bumps around 20 and 60 mesh widths.
Other snapshots of Run~2 however do not show this feature. 
The bumps in the structure functions indicate the presence of
preferred scale sizes in the simulation. 
This means that even if the starting point is a fully
isotropic hydrodynamic solution of supersonic turbulence, then
preferred scales can be encountered by selecting shocked regions
that have a typical filamentary length of some hundred mesh widths (see Fig.~\ref{fig:shock}).

\section{Conclusion}
\label{sec:conclusion}

We have used observational data of shocked H$_2$ emission in OMC1 to
show that structure 
functions at scales 70 -- 3$\cdot 10^4\,$AU (3.4$\cdot 10^{-4}$ --
0.15\,pc) exhibit unusual 
scaling exponents for $p>3$. The scaling exponents are nearly constant
for $p>3$ and smaller than predicted by both \cite{she1994}
and \cite{boldyrev2002a}. 

In three simulations we have selected shocked regions by
imposing requirements on the value of the velocity divergence,
$\vec\nabla\cdot\vec{u}$.
In certain important respects the simulations presented here are 
then remarkably successful in reproducing the statistical behaviour observed 
in OMC1. In other equally important areas, they fail. Let us first 
reiterate the success.

We have
found that by only including shocks that are 
relatively strong
($D_0 < -1.5$), the unusual scaling exponents of the
observations are reproduced in the simulations. By contrast, a
scaling following that of \cite{she1994} or \cite{boldyrev2002a} is found
when all points in the simulations are included in the data.
An explanation for this behaviour is as follows.
Enhanced energy content at small scales, relative to larger scales,
implies smaller slopes, that is, smaller values of $\zeta_p$. Both the
observational data of OMC1, and some of those subsets of the
simulations selected only to include shocks, show that the values of
$\zeta_p$ are reduced for $p\ge4$.
Since structure functions of high order $p$ are dominated by regions of strong
velocity differences, it follows that the observed excess of small
scale energy is associated with regions of large velocity differences.
These are likely to be the regions of strong shocks, as is evidenced
by the fact that reduced values of $\zeta_p$ are most clearly seen in
subsets of the simulations that select the most strongly convergent
high density regions.  

The present work does not however furnish any explanation of why
departure from the She-Leveque or Boldyrev scaling occurs at the
specific value of
$p\ge4$. It is possible that the critical value of $p$ is in some way
connected with the physical nature of the shocks, for example the fact
that they are smoothed in the simulations, mimicking the structure of
continuous (C-) type shocks, rather than jump (J-) type shocks
\citep[][ and references therein]{flower2003}.

We now turn to the failure of the simulations.
The structure functions of the observations in OMC1 all deviate from
power laws and exhibit
clear bumps
around $10^3$AU, exemplified by the third order structure function in
Fig.~\ref{fig:scal_obs}. 
This cannot in general be reproduced by the
simulations.
There appears to be two possible explanations for this observed
behaviour. The first is that the deviation from power laws is due to
protostar formation and associated outflows at a preferred scale. The
second is that the behaviour is in some way inherent in the nature of
the turbulence as opposed to the presence of
protostars. 

Turning to the first suggestion, the process of star formation pulls
structure of a certain size out of the 
cascade and creates outflows, injecting energy into a turbulent
background. Gravitational energy and angular 
momentum is spewed out of the star via such outflows and turned into local
 turbulence, hence increasing the overall turbulent content of the gas
 -- and restarting the whole cascade process.
Such outflows are of course not present in the simulations.

The second suggestion, that the deviation from power law is somehow
inherent in the nature of the turbulence, requires that there is some
non-statistical element in this medium which is otherwise governed by
statistical considerations. This may arise through our selection of
strong shocks as a subset of the whole. 
We have seen in Fig.\ref{fig:sim_512g} that traces of bumps in the structure
functions are found in 
one snapshot of Run~2 when highly shocked material is selected. The
deviations of the 
structure functions from power laws are not as pronounced in
the simulation as in the observations of OMC1. However this finding
provides some evidence
that part of the explanation for the deviations from power laws of the
structure functions in OMC1 arises from the fact that we observe
preferentially shocked gas. 
As departures from power law behaviour are only evident in a single
snapshot and 
not throughout the simulation at other times, this suggestion remains
tentative. 

In order to explore the reasons for the departure from power law
behaviour, more advanced simulations are necessary. These should
include self-gravity and energy feedback from protostellar zones
through outflows and should ultimately incorporate ionization and
magnetic fields.

\begin{acknowledgements}

We thank {\AA}ke Nordlund for simulating discussions and advice in
defining this project.
DF and MG would like to acknowledge the support of the Aarhus Centre
for Atomic Physics (ACAP), funded by the Danish Basic Research
Foundation and the Instrument Center for Danish
Astrophysics (IDA), funded by the Danish Natural Science Research
Council.
We would also like to thank the Directors and Staff of the CFHT for
making possible the observations used in this paper.
The Danish Center for Scientific Computing is acknowledged for granting
time on the Horseshoe cluster in Odense.

\end{acknowledgements}

\appendix
\section{The forcing function}
\label{ForcingFunction}

For completeness we specify here the forcing function used in the
present paper.
It is defined as \citep{brandenburg2001}
\begin{equation}
\vec{f}(\vec{x},t)={\rm Re}\{N\vec{f}_{\vec{k}(t)}
\exp[{\rm i}\vec{k}(t)\cdot\vec{x}+{\rm i}\phi(t)]\},
\end{equation}
where $\vec{x}$ is the position vector.
The wavevector $\vec{k}(t)$ and the random phase
$-\pi<\phi(t)\le\pi$ change at every time step, so $\vec{f}(\vec{x},t)$ is
$\delta$-correlated in time.
For the time-integrated forcing function to be independent
of the length of the time step $\delta t$, the normalization factor $N$
has to be proportional to $\delta t^{-1/2}$.
On dimensional grounds it is chosen to be
$N=f_0 c_{\rm s}(|\vec{k}|c_{\rm s}/\delta t)^{1/2}$, where $f_0$ is a
nondimensional forcing amplitude.
At each timestep we select randomly one of many possible wavevectors
with length between 1 and 2 times the minimum wavenumber in the box, $k_1$.
The average wavenumber is $k_{\rm f}=1.5k_1$.
We force the system with transverse non-helical waves,
\begin{equation}
\vec{f}_{\vec{k}}=
\left(\vec{k}\times\vec{e}\right)/\sqrt{\vec{k}^2-(\vec{k}\cdot\vec{e})^2},
\label{nohel_forcing}
\end{equation}
where $\vec{e}$ is an arbitrary unit vector
not aligned with $\vec{k}$; note that $|\vec{f}_{\vec{k}}|^2=1$.

\bibliographystyle{aa}
\bibliography{bibliography}

\end{document}